\documentclass[reprint,amsmath,amssymb]{revtex4-1}
\usepackage[utf8x]{inputenc}
\usepackage{graphicx}
\usepackage{dcolumn}
\usepackage{bm}
\usepackage{color} 
\usepackage{amsmath}
\usepackage{tensor}

\usepackage{soul} 
\usepackage[normalem]{ulem} 

\begin{document}

\preprint{APS/123-QED}


\title{Fibonacci number systems and the localization criterion in the many-body Aubry-Andr\'{e} model}

\author{Bal\'azs Het\'enyi$^{1,2,3,*}$}
\affiliation{$^1$Department of Theoretical Physics, Budapest University of Technology and
  Economics, H-1111 Budapest, Hungary \\ and \\
  $^2$ MTA-BME Lend\"{u}let "Momentum"  Open Quantum Systems Research Group, Institute of Physics,
 Budapest University of Technology and Economics, M\H{u}egyetem rkp. 3, H-1111 Budapest, Hungary 
 \\ and \\
 $^3$Institute for Solid State Physics and Optics, HUN-REN Wigner Research Centre for Physics,  H-1525 Budapest, P. O. Box 49, Hungary \\ 
 $^*$ Corresponding author, email: hetenyi.balazs@ttk.bme.hu}

\date{\today}
\begin{abstract}
We introduce an extension of the Fibonacci number system, we call the fractional Fibonacci number system, which interpolates between the Fibonacci number system (used for natural numbers) and the irrational base-$\phi$  number system, which can be used to represent real numbers.   The number system finds its use in interpreting the localization phase diagram of the many-body non-interacting Aubry-Andr\'{e} model.   For finite system sizes a localization criterion can be obtained if the particle density is written in the fractional Fibonacci number system.  In the thermodynamic limit, the criterion remains, but in this case the particle density is expressed in base-$\phi$.   The nature of the thermodynamic limit is also discussed.
\end{abstract}

\maketitle


\label{sec:intro}

{\it Introduction.-}    The delocalization-localization transition (DLT) occuring in disordered~\cite{Abrahams79,Langedijk09,Evers08} and quasiperiodic~\cite{Aubry80,Martinez18,Dominguez-Castro19} systems is a topic of ongoing current interest.  In the context of the latter the quintessential one-dimensional lattice model is the Aubry-Andr\'{e}-Harper model (AAM)~\cite{Aubry80}, consisting of a nearest neighbor hopping term ($t$) and a modulating potential of strength $W$.  The potential also depends on an irrational parameter, $\alpha$, usually taken to be the golden ratio.  The AAM is equivalent to the Harper model~\cite{Harper55} used to describe the quantum Hall effect~\cite{vonKlitzing80,Tsui82,Thouless82}.  The AAM motivated many mathematical studies~\cite{Jitomirskaya99,Avila06,Avila09,Avila23} which investigated the characteristics of single particle eigenvalues and eigenstates.  Jitomirskaya~\cite{Jitomirskaya99} showed that single-particle states localize at a finite interaction strength.   Avila et al.~\cite{Avila09} showed that the single-particle energy spectrum of the AAM  is a Cantor set.   A very recent study~\cite{Avila23} showed that commensurate systems exhibit gaps (when $\alpha$ is approximated as a rational number) which survive in the incommensurate limit (when the irrational limit for $\alpha$ is taken).   In mathematical circles this set of problems is referred to as the "Ten Martini Problem" or "Dry Ten Martini Problem"~\cite{Jitomirskaya99,Avila06,Avila09,Avila23}.   This is for historical reasons, and certainly not due to any excessive prevalence of alcoholism among practitioners of the field.  We mention in passing that the AAM has a very long history, including realization in experimental settings~\cite{Billy08,Roati08,Modugno09,Kohlert19} as well as theoretical studies of the original model and extensions~\cite{Johansson91,Biddle10,Biddle11,Ganeshan13,Ganeshan15,Bistritzer11,Monthus17,Li20,Padhan22,Goncalves23a,Goncalves23b,Dziurawiec24,Chi24,Zhang25,Goswami25,Lu25,Gandhi25,Sahoo26,Zhang26,Liu26,Bonsel26,Jeon26}.\\

In this work we will study the non-interacting many-body version~\cite{Cookmeyer20,Varma15,Hetenyi24,Hetenyi25,Mastropietro15,Xu19,Huang24} of the AAM.   Recent studies~\cite{Cookmeyer20,Hetenyi25} have shown an unusual density ($\rho$) vs. potential strength ($W$) phase diagram in this case.  When extrapolating to the thermodynamic limit, rational densities exhibit a phase transition at $W=2t$, just like the single particle states, but certain irrational densities show a $W=0$ transition.    In this study, we will use the Fibonacci base number system, discovered in 1957 by George Bergman~\cite{Bergman57}, to explain these results.\\

Extending a previous study~\cite{Hetenyi25} we will show that there are three relevant categories of numbers in the interpretation of the $\rho$ vs. $W$ phase diagram.  On the one hand, at rational densities, as mentioned before, the AAM undergoes a localization transition at $W=2t$.  The irrational numbers divide into two further categories.  One of these is formed by the set of irrational numbers which can be produced by taking the infinite limit of Fibonacci ratios or finite sums thereof.  At these fillings, the localization transition occurs at $W=0$.   For all irrational numbers outside this category the transition occurs at $W=2t$, as it does for rational fillings.  We also show that this classification can be made more transparent using an extension of the Fibonacci number system, which we will call the fractional Fibonacci number system.  The Fibonacci number system represents positive integers, whereas any real number can be represented in the base-$\phi$ number system.  The fractional Fibonacci number system we introduce interpolates between the two cases.  The fractional Fibonacci number system can represent any fraction of the form $M/F_n$, where $M$ is an integer and $F_n$ is a Fibonacci number.  As $n \rightarrow \infty$  the range of numbers which can be represented becomes the entire real axis and the base-$\phi$ number system is recovered.  We also demonstrate the use of this number system in defining a localization criterion for the many-body non-interacting Aubry-Andr\'{e} model.   If a density for some system size $L=F_n$ is written in the appropriate fractional Fibonacci number system, and the number does not change as the system size (and the fractional Fibonacci number system) is changed, the system at that density will be localized for all finite potential strength.  At other densities, a genuine metal-insulator transition occurs at $W=2t$.
\\


{\it Model and three numerical example calculations.-}. The Hamiltonian of the Aubry-Andr\'{e} model is given by,
\begin{equation}
H = \sum_{j=1}^L \left[ (-t)(c_j^\dagger c_{j+1} + c_{j+1}^\dagger c_j) + W \cos (2 \pi \alpha j) n_j \right],
\end{equation}
where $t$ denotes the hopping strength, $W$ denotes the potential strength, $\alpha$ denotes the golden ratio (approximated here as a ratio of Fibonacci numbers).  $c_j$($c_j^\dagger$) are fermionic annihilation(creation) operators.  Throughout this study, periodic boundary conditions are assumed.   The main focus of our study is the many-body version of this model: we will study systems of size $L$ and finite particle number $N$, the density is then defined as $\rho = N/L$.  The golden ratio will be approximated as $\alpha = F_{n+1}/F_n$, and the system size will always correspond to $L=F_n$.\\

We diagonalize the AAM Hamiltonian under periodic boundary conditions, resulting in a set of states on the lattice, $\Phi_\lambda(j)$, where $\lambda$ denotes the state index, and $j$ denotes the lattice site.  For a system with $N$ particles, the ground state wave function is a Slater determinant of the $N$ lowest energy states,
\begin{equation}
\Psi(j_1,...,j_N) = \mbox{Det} \left[ \Phi_\lambda(j_\mu) \right]; \lambda = 1,...,N.
\end{equation}
We calculate the variance, by obtaining $Z_1$, defined in the case of  band systems as
\begin{equation}
Z_1  =  \mbox{Det} \left[ U_{\lambda \lambda'}^{(1)} \right],
\end{equation}
where
\begin{equation}
U_{\lambda \lambda'}^{(1)}= \sum_{j=1}^L \phi^*_\lambda(j) \exp\left( i \frac{2 \pi}{L}j\right) \phi_{\lambda'}(j).
\end{equation} 
The variance~\cite{Hetenyi24,Hetenyi25,Resta98,Resta99} (normalized and centered second moment) in our calculations takes the form,
\begin{equation}
M_2 = \frac{L^2}{2\pi^2 N} (1 - |Z_1|).
\end{equation}

\begin{figure}[ht]
 \centering
 \includegraphics[width=8.5cm,keepaspectratio=true]{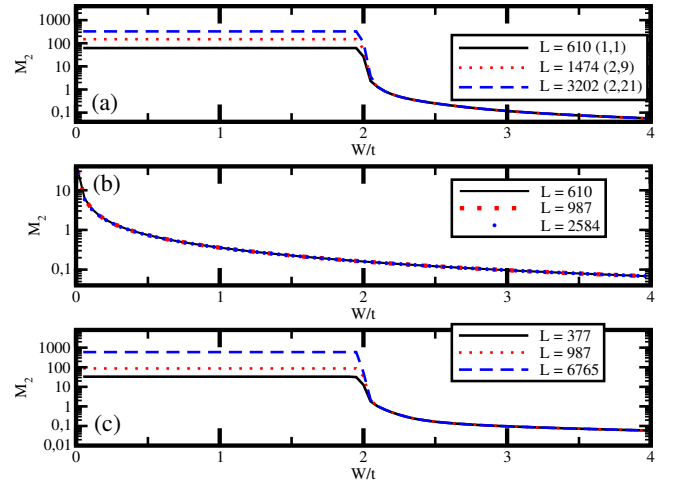}
 \caption{Numerical results for the variance (centered second moment) of the polarization ($M_2$) as a function of the potential strength ($W/t$) for three particle densities (a) $\rho = \frac{1}{2}$, (b) $\rho = \frac{2}{1 + \sqrt{5}}$, (c) $\rho = \frac{1}{\sqrt{3}}$.   The numbers in parenthesis in panel (a) indicate the first two numbers in the sequences used to generate the system size according to the rule, $F_{n+1} = F_n + F_{n-1}$.  In panel (a) the density is exact, in panels (b) and (c) we make rational approximations to the irrational densities.  }
 \label{fig:M2three}
\end{figure}

To motivate this study, we first present a set of numerical calculations.   Fig. \ref{fig:M2three} shows the results of the centered second moment (variance) of the polarization~\cite{Hetenyi24,Hetenyi25,Resta98,Resta99} divided by the number of particles as a function of $W/t$ for three particle densities: one a rational number ($\rho = \frac{1}{2}$), one an irrational number ($\rho = \frac{2}{1 + \sqrt{5}}$, approximated as $\frac{F_{n-1}}{F_n}$), and one an irrational number unrelated to the golden ratio ($\rho = \frac{1}{\sqrt{3}}$, approximated as the fraction $\rho = m/L$ closest to $\frac{1}{\sqrt{3}}$).   In some cases we used different series~\cite{Vajda89}, not the usual Fibonacci (whose first two elements are $1,1$), but series generated by the rule $F_{n+1} = F_n + F_{n-1}$ but with a different pair of starting numbers.   Such series still produce the golden section in the infinite $n$ limit, therefore the AAM exhibits identical properties to the one in which the usual Fibonacci sequence is used.  For each of these densities three different system sizes are shown.  For the first and third cases significant size effects are found in the region $W/t<2$, but for larger $W/t$ the curves fall on each other, the system size dependence disappears.   This is a clear indication of a metal-insulator transition.  One can calculate the size scaling exponent and it is found that it takes the value of unity for $W/t<2$ (metallic phase), and zero for $W/t>2$ (insulating phase).   The second case behaves differently, all three curves fall on each other, meaning that the size scaling exponent is zero for the entire range of $W/t$, and that the system is always insulating for finite $W/t$.  These are example calculations, but the results were found to be general.  We remark that the first two cases (rational and golden ratio irrational) were studied for larger system sizes in Ref. \cite{Hetenyi25}.    \\

{\it Zeckendorf theorem}.-  The Zeckendorf theorem~\cite{Zeckendorf72,Samons94} is a helpful device in interpreting the above results.  According to this theorem all natural numbers suffer a unique decomposition in terms of a sum of Fibonacci numbers (subject to certain constraints, for example, no two consecutive numbers can appear).   The procedure to arrive at the representation of some natural number $N$ is based on a sequential greedy algorithm.  One finds the largest Fibonacci number, $F_n$ satisfying $F_n \leq N$.   $F_n$ is the first member of the representation.  One then subtracts $F_n$ from $N$, and continues to look for the largest Fibonacci number in the remainder.  \\

The decomposition of the three different densities shown in Fig. \ref{fig:M2three} for different system sizes can be instructive.  For $\rho = 1/2$ let us take the three even system sizes $L=34,144,610$ (particle numbers $N=17,72,305$, respectively).   We find,
\begin{eqnarray}
\label{eqn:half}
\rho_{34} &=& \frac{1+3+13}{34}, \\ \nonumber \rho_{144} &=& \frac{1+3+13+55}{144}, \\ \nonumber  \rho_{610} &=& \frac{1+3+13+55+233}{610},
\end{eqnarray} 
where the subscripts indicate the particular system size.  We see that as the system size increases, the number of terms in the numerator (the result of the Zeckendorf decomposition) increase.   For the second density considered in Fig. \ref{fig:M2three}, the Zeckendorf decomposition always gives a single term.  Considering again the system sizes $L=233,377,610$ ($N=144,233,377$, respectively).  In this case all three densities can be written as,
\begin{equation}
\rho = \frac{F_{n-1}}{F_n},
\end{equation}
with $n = 13,14,15$.   Not that $\lim_{n \rightarrow \infty} \frac{F_{n-1}}{F_n} \rightarrow \frac{2}{1 + \sqrt{5}}$, in other words, increasing $n$ extrapolates to the desired density.   A crucial difference between this case and the previous one is that the numerator (the Zeckendorf decomposition) consists of a single term.  As it was found in Refs. \cite{Cookmeyer20,Hetenyi25} densities consisting of numerators with a finite number of terms exhibit only an insulating phase.  For completeness we consider the third case, $\rho = \frac{1}{\sqrt{3}}$ for the system sizes $L=233,377,4181$ ($N=134,218,2413$, respectively).  In these cases,
\begin{eqnarray}
\rho_{233} &=& \frac{3+8+34+89}{233}, \\ \nonumber \rho_{377} &=& \frac{1+5+13+55+144}{377}, \\ \nonumber  \rho_{4181} &=& \frac{2+5+55+144+610+1597}{4181},
\end{eqnarray} 
where the subscripts indicate the system sizes.  Again, the number of terms in the sum increase as $L$ is increased. \\

{\it Fibonacci number system, fractional Fibonacci number system, and base-$\phi$ numbers}.-  Based on the Zeckendorf decomposition one can construct a binary number system to represent natural numbers.   To write a natural number $N$ in the Fibonacci base, one first carries out the Zeckedorf decomposition.   If the Fibonacci number $F_n$ appears then the $n$th digit of $N$ in the Zeckendorf number system is unity, otherwise it is zero.   For example, the number $N=12=1+3+8=F_2+F_4+F_6=10101_F$, where the subscript $F$ indicates that $N$ is expressed in the Fibonacci number system.  Note that the coefficient of $F_2=1$ corresponds to the first digit.  In general, a natural number $N$ can be written,
\begin{equation}
N = \sum_{j = 2}^\infty a_j F_j,
\end{equation}   
with $a_j \in [0,1]$ and with the restriction that no consecutive coefficients can be unity.\\

The base-$\phi$ number system~\cite{Bergman57} is a related number sytem whose base is the irrational number $\phi = \frac{1+\sqrt{5}}{2}$.  A base-$\phi$ number also has digits of zero or unity and is of the general form, $\sum_{j \in \mathbb{Z}} a_j \phi^j$ with $a_j = 0,1$.   As in the Fibonacci system, no neighboring digits of unity are allowed.  \\

The Fibonacci number system represents only integers, not numbers between zero and unity.  This is definitely an issue, since we would like to represent the particle density, $\rho = N/L$.  In theory it is possible to use the reciprocals of Fibonacci numbers to represent fractions, but in that case the Zeckendorf representation becomes inapplicable.   The base-$\phi$ number system can represent fractions as well, but the connection with the Zeckendorf decomposition is not immediate.\\

These questions can be addressed by constructing an intermediate number system between the Fibonacci and the base-$\phi$ number system, which we will refer to as the fractional Fibonacci number system.  The essential point is that in our case the system size $L=F_n$ is a Fibonacci number.   In addition, for a given system size, one does not have to represent all fractions, only those of the form, $N/L$, so the division required is one with $F_n$.   To see how our extended representation is constructed, we first provide a simple example.  Consider the number base-$10$ number $223$.   The largest decimal number that occurs if this numbers is broken into components is $10^2$, occurring two times.  If we divide this number by the next largest base number, $10^3$, the number simply shifts to the right, beyond the decimal point, and it becomes $0.223$.  Our extended Fibonacci base construction is analogous.   For a number, $\rho = N/L$, where $L=F_n$, a Fibonacci number,  we first decompose $N$ according to Zeckendorf.  In this decomposition, the largest Fibonacci number, that can appear is $F_{n-1}$.   It can appear with a coefficient of zero or unity.  We now divide each term in the Zeckendorf decomposition by $F_n$ to obtain the density, $\rho$.   This will shift all digits beyond the radix point, as $10^3$ did for the number $223$.  The new number will represent fractions up to resolution $F_n^{-1}$.   
Note that the scheme generalizes readily to numbers larger than unity, the only restriction is that the number, expressed as a fraction, should have $F_n$ as the denominator.\\

Let us consider an example.  Suppose that the system size is $L=F_6=8$, and $N=6$.  The Zeckendorf decomposition for $N$ gives $N=F_2+F_5$, which in the Fibonacci number system is $N = 1001_F$.  According to our scheme, dividing by $F_6$ shifts digits, giving,
\begin{equation}
\frac{N}{L} = \frac{6}{8} = 0.1001_{F_6}.
\end{equation}
We introduced the notation $F_6$ as a subscript for the fractional Fibonacci number system for fractions of the form $M/F_6$ ($M \in \mathbb{Z}$).\\

In the $F_n$ number system the base numbers start at $F_2/F_n$, whose coefficent is the digit farthest from the radix point.  The coefficient of $F_{n-1}/F_n$ gives the digit just after the radix point.  A number of the form $\omega = M/F_n$ (possibly larger than one), where $M$ is a natural number can be written as,
\begin{equation}
\omega = \sum_{j=2}^\infty c_j \frac{F_j}{F_n},
\end{equation}
where $c_j \in [0,1]$, and no consecutive coefficients can be unity.  The coefficients, $c_j$ can be determined from the Zeckendorf decomposition of $M$. \\

In general, for a system with finite system size $L=F_n$ and density $\rho= N/L$, the Zeckendorf decomposition gives a sum in which the highest Fibonacci index is $n-1$,
\begin{equation}
\rho = \sum_{j=2}^{n-1} a_j \frac{F_j}{F_n},
\end{equation}
where, again, $a_j\in [0,1]$, and no consecutive indices can be unity.   The thermodynamic limit can be taken by making $L=F_n$ (or $n$) large. Taking advantage of the fact, that
\begin{equation}
\label{eqn:FperF}
\lim_{m \rightarrow \infty} \frac{F_{j+m}}{F_{n+m}} = \phi^{j-n},
\end{equation}
This equation is proven in the Appendix.  We write an approximate particle density as,
\begin{equation}
\rho \approx \sum_{j=2}^{n-1} a_j \phi^{j-n}.
\end{equation}
This sum can be rewritten as,
\begin{equation}
\rho \approx \sum_{j=1}^{n-2} b_j \phi^{-j},
\end{equation}
with $b_1 = a_{n-1},b_2 = a_{n-2},...,b_{n-2} = a_2$.   Note that the change in coefficients does not effect the restriction that consecutive coefficients can not be unity, since the new coefficients, $b_j$, are the same as $a_j$, only their order is inverted.  In the thermodynamic limit, the density becomes,
\begin{equation}
\label{eqn:rhoexact}
\rho = \sum_{j=1}^\infty b_j \phi^{-j},
\end{equation}
in other words, the base-$\phi$ is recovered.  This last expression is an exact expression for the density.  It is also proven in the Appendix.  \\

We note in passing that base-$\phi$ numbers form an iterative function system (self-similarity).   We can write the set of all base-$\phi$ numbers in the interval between zero and one as,
\begin{equation}
S = \left\{ \sum_{j=1}^\infty b_j \phi^{-j}: b_j \in 0,1 ; \mbox{no consecutive 1s}\right\}.
\end{equation}
The iterative structure is imposed by the "no consecutive 1s" rule.  The set $S$ can be broken into two nonoverlapping subsets depending whether the first digit after the radix is zero or one.   We label these two subsets as $S_0$ and $S_1$, respectively.    To obtain $S_0$, we set the first coefficient, $b_1$ to zero, resulting in 
\begin{equation}
S_0 =  \left\{ \sum_{j=2}^\infty b_j \phi^{-j}: b_j \in 0,1 ; \mbox{no consecutive 1s}\right\}.
\end{equation}
A simple shift in indices gives,
\begin{equation}
S_0 = \phi^{-1} S.
\end{equation}
To obtain $S_1$, we set $b_1 = 1$, which implies that $b_2=0$, because of the no consecutive ones rule.  This results in,
\begin{equation}
S_1 =  \left\{  \phi^{-1} + \sum_{j=3}^\infty b_j \phi^{-j}: b_j \in 0,1 ; \mbox{no consecutive 1s}\right\}.
\end{equation}
Again, a simple shift of indices gives
\begin{equation}
S_1 = \phi^{-1} + \phi^{-2} S.
\end{equation}
$S$ can be formed from the union $S = S_0 \cup S_1$, making the self-similarity of $S$ manifest, since both $S_0$ and $S_1$ depend on $S$. \\

The criterion for localization for a given density can be seen from the fractional Fibonacci number system.   In table \ref{tab:Fn} three example densities, $\rho = F_{n-1}/F_n$ and $\rho = 1/2$, and $\rho = 1/\sqrt{3}$, one from each category, are analyzed for three system sizes, $L = F_6, F_{12}, F_{15}$ as well as the thermodynamic limit.  Each density is represented in the fractional Fibonacci number system of the corresponding system size $L = F_n$.  For the density at which no genuine metal-insulator transition takes place (only at $W=0$), namely $\rho = F_{n-1}/F_n$, the number expressing the density does not change in the different representations, including the base-$\phi$.  The number is always $\rho = 0.1_{F_n}$, irrespective of the value of $n$.  For $\rho=1/2$, a metal-insulator transition occurs at $W=2t$.  For this density, when written in the fractional Fibonacci base representation, the number of nonzero digits increase with the system size.  Based on the pattern one can extrapolate that in the thermodynamic limit the density in base-$\phi$ is $0.\overline{010}_\phi$.  For $\rho = 1/\sqrt{3}$ the number of nonzero digits increases with system size, but there is no definite pattern, as it happens for $\rho=1/2$. \\

\begin{widetext}
\begin{table}[h]
    \centering
    \begin{tabular}{|l|c|c|r|}
        \hline
               System size & $\rho = F_{n-1}/F_n$ & $\rho = 1/2$ & $\rho = 1/\sqrt{3}$ \\
        \hline \hline
        $F_6$ & 0.1$_{F_6}$ & 0.0101$_{F_6}$ & 0.1$_{F_6}$\\ \hline
        $F_{12}$ & 0.1$_{F_{12}}$ & 0.0100100101$_{F_{12}}$ & 0.0101001001$_{F_{12}}$ \\ \hline
        $F_{15}$ & 0.1$_{F_{15}}$ & 0.0100100100101$_{F_{15}}$ & 0.0101001010001$_{F_{15}}$ \\ \hline
        $F_\infty$ (thermodynamic limit) & 0.1$_\phi$ & 0.$\overline{010}_\phi$ & 0.0101001010000101010001...$_\phi$ \\
        \hline
    \end{tabular}
    \caption{Example table}
    \label{tab:Fn}
\end{table}
\end{widetext}

{\it The thermodynamic limit}.-  The previous criterion loses its strict meaning in the thermodynamic limit, since it is defined using a finite system as a starting point.  For a system size $L=F_n$ the densities, $\rho = N/L$, with $N=1,...,F_n-1$, in base-$F_n$ will all have at most $F_n-1$ digits.  Increasing the system size to $F_{n+1}$ will generate new densities with larger numbers of digits, but the densities "inherited" from the $F_n$ system will remain with the same number of digits.  To clarify this, let us take two small consecutive system sizes, $L=5$ and $L=8$, and express all densities in the$F_4$ and $F_5$ number systems, respectively.  For the former, the four possible densities are $0.001_{F_4}$, $0.001_{F_4}, 0.01_{F_4}, 0.1_{F_4}, 0.101_{F_4}$, corresponding to $1/5,2/5,3/5,4/5$, respectively.  The larger system size, $L=8$ will "inherit" some of the the densities from $L=5$ (by way of a shift of indices in the Zeckendorf decomposition in each case).  These inherited densities are $\rho = 2/8,3/8,5/8,7/8$.  These will, again, have the same digits in the same places in the $F_5$ representation.  The remaining densities, $\rho=1/8,4/8,6/8$ can be written $0.0001_{F_5}, 0.0101_{F_5}, 0.1001_{F_5}$, respectively.  They are the inherited densities with one extra digit of unity added such that the Zeckendorf constraint is satisfied.  As the system size is increased further, the original densities will tend to localize, while the newly generated ones will show a finite metal-insulator transition at $W=2t$.   However, in the above argument, $L=5$ was an arbitrary starting point.  

For large $L$, when the thermodynamic limit is reached, and the base-$\phi$ representation is recovered, one can extrapolate the above example as follows.  On can compare a large system to an even larger one and check whether the density, expressed in the base-$\phi$ representation, remains the same, or acquires new digits.   For example, the density $\rho = 1/L$ will always have a transition at $W=2t$.   Another way to think about it is whether in the thermodynamic limit $\rho$ expressed in base-$\phi$ is intensive or depends on the extent $L$.\\ 

A further difficulty arises in the thermodynamic limit when one considers the following densities, 
\begin{eqnarray}
\label{eqn:gpm}
\gamma_+ = \lim_{n \rightarrow \infty} \frac{F_{n-1} + 1}{F_n}, \\
\nonumber
\gamma_- = \lim_{n \rightarrow \infty} \frac{F_{n-1} - 1}{F_n}.
\end{eqnarray}
For a finite value of $n$ the numerator of $\gamma_+$ gives the value $10....01_F$, where the number of zeros between the ones is $n-1$.  One can convert this number to base-$\phi$ obtaining an approximation of $\gamma_+ \approx 0.10....01_\phi$.   As the limit $n\rightarrow \infty$ is taken, the base-$\phi$ number becomes $\gamma_+ = 0.1_\phi $.  $\gamma_+$ tends to $\gamma$ from above.   Also instructive is the behavior of $\gamma_-$.  After trying a few examples, we see that finite $n$ give rise to $0.0101...01_\phi$ numbers with finite digits, but the pattern $01$ repeating.  As $n\rightarrow \infty$ $\gamma_-$ tends to $0.\overline{01}_\phi$.   But there is an important relation between $\gamma = 0.1_\phi$ and $\gamma_- = 0.\overline{01}$.  The connection between the two numbers is the analog of that between the base ten number, $0.2$ and the number $0.1\overline{9}$.  These two numbers are equal.  Similarly, when the limit $n \rightarrow \infty$ is taken, $\gamma = \gamma_+ = \gamma_-$.   Defining an interval between $\gamma_+$ and $\gamma_-$ is finite for finite $n$, but zooms in on $\gamma$ as $n$ tends to infinity. \\

\begin{figure}[ht]
 \centering
 \includegraphics[width=8.5cm,keepaspectratio=true]{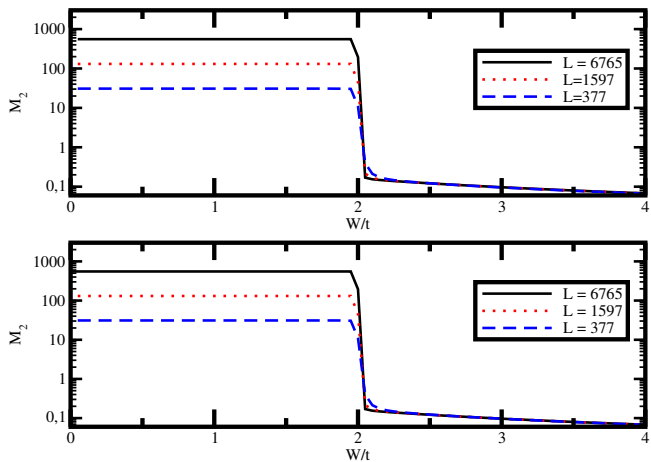}
 \caption{Centered variance of the polarization for (a) $\gamma_+$ and (b) $\gamma_-$ (see Eq. (\ref{eqn:gpm})).  $\gamma_+$ and $\gamma_-$ differ from $\frac{F_{n-1}}{F_n}$ by plus one or minus one (respectively) particle.  Clear evidence is found for a transition at $W=2t$.  The results for $\gamma$ are shown in Fig. \ref{fig:M2three}, panel (b).}
 \label{fig:M2two}
\end{figure}

Calculations for $\gamma_+$ and $\gamma_-$ are shown in Fig. \ref{fig:M2two} for three system sizes each.  In contrast to $\gamma$, for which calculations are shown in Fig. \ref{fig:M2three}, panel (b), there is a definite transition at $W=2t$, even though, in the thermodynamic limit, the three points, $\gamma_+, \gamma_-$, and $\gamma$ coalesce.  The localization behavior for $\gamma$ depends on how the thermodynamic limit is taken: a particle density of $0.1_\phi$ leads to a localized phase for any $W/t$, while approaching this number from above or below leads to a metal-insulator transition at finite $W/t$. \\

{\it Conclusion}.-  We investigated the use of base-$\phi$ numbers in interpreting the localization transition in the many-body Aubry-Andr\'{e} model with a modulation given by the golden section.  When the particle density is written in base-$\phi$ the dependence of the number of finite digits after the radix point determines where the localization transition occurs.  In base-$\phi$ the set of irrational numbers which are the limits of Fibonacci ratios and sums thereof are easily distinguished, because other numbers are fractions with infinite digits.  On the other hand, in the thermodynamic limit, when the system size is infinite, this classification loses its meaning.  It was also shown that in this limit the localization transition for a given filling can depend on the direction from which it is approached. \\

\section*{Acknowledgements} The author acknowledges helpful discussions with Erg\"un Yalc\i n.  The author gratefully acknowledges support by HUN-REN 3410107 (HUN-REN-BME-BCE Quantum Technology Research Group), by the National Research, Development and Innovation Fund of Hungary within the Quantum Technology National Excellence Program (Project No. 2017-1.2.1-NKP-2017-00001), by Grants No. K142179, No. K142652, and No. FK142601 and by the BME-Nanotechnology FIKP Grant No. (BME FIKP-NAT). \\

\section{Appendix: Three relevant proofs}

In this Appendix, we provide the steps of the prove three statements used in the text.  The first statement to be proven is Eq. (\ref{eqn:FperF}) itself.  The second and third are both connected to Eq. (\ref{eqn:rhoexact}) and they are stated below.   \\

To prove Eq. (\ref{eqn:FperF}), our starting point is the expression,
\begin{equation}
\label{eqn:Fnphiphi}
F_n = \frac{\phi_+^n - \phi_-^n}{\phi_+ - \phi_-},
\end{equation}
where,
\begin{equation}
\phi_\pm = \frac{1 \pm \sqrt{5}}{2}.
\end{equation}
$\phi_+$ and $\phi_-$ are the two roots of the infinite limit Fibonacci equation, $x^2 - x - 1 = 0$.  $\phi_+$ is known as the golden section, and is denoted $\phi$ in the main text, where $\phi_-$ is not needed.   Eq. (\ref{eqn:Fnphiphi}) can be proven by showing it to be true for $n=1$, and applying induction.  One can then prove Eq. (\ref{eqn:FperF}) using the fact that $\phi_+ > \phi_-$ and setting $\phi_+ = \phi$. \\

Regardig Eq. (\ref{eqn:rhoexact}), we make two statements.  One is that any real number $R$ can be expressed as,
\begin{equation}
\label{eqn:Rbphi}
R = \sum_{j \in \mathbb{Z}}^ \infty b_j \phi^j, 
\end{equation}
such that $b_n \in [0,1]$ and there can be no consecutive coefficients which are unity.  To show this, one starts with the Fibonacci identity,
\begin{equation}
\phi = 1 + \frac{1}{\phi}.
\end{equation}
multiplied by $\phi^n$,
\begin{equation}
\label{eqn:phin}
\phi^{n+1} = \phi^n + \phi^{n-1}.
\end{equation}
To show that $R$ can be expanded in base $\phi$, (Eq. (\ref{eqn:Rbphi})), according to the conditions given on the coefficeints $b_j$, one can apply the Zeckendorf decomposition, using not the Fibonacci numbers, but powers or $\phi$.  We proceed through an iterative greedy algorithm.   We first look for the largest $\phi^n$ smaller than or equal to $R$.  If $R=\phi^n$ for some integer $n$, we are done.  If not, then we subtract $\phi^n$ from $R$, and we note that $b_n = 1$, and continue until we obtain zero.  The reason that the Zeckendorf constraints on the coefficients will be satisfied can be seen by writing out the iteration steps of the sequential greedy algorithm as follows.   The condition to determine $\phi^n$ can be expressed as,
\begin{equation}
\phi^n < R < \phi^{n+1}.
\end{equation}
We can subtract $\phi^n$ from all three "sides" resulting in,
\begin{equation}
0 < R - \phi^n < \phi^{n+1} - \phi^n = \phi^{n-1},
\end{equation}
where in the last part, Eq. (\ref{eqn:phin}) was used.  Since the starting number for the next iteration, $R - \phi^n$ is smaller than $\phi^{n-1}$, $b_{n-1}$ can not be a coefficient with value of unity, if $b_n$ is.  This proves that any real number, $R$ can be written in base-$\phi$.  For our purposes the density is $0 \leq \rho \leq 1$, so $\rho$ can always be represented in base-$\phi$. \\

The second statement we make is a formal one whose purpose is to complete the connection between the intermediate Fibonacci number system used in the text, and its thermodynamic limit.  Let $A$ denote the set of numbers of the form $N/L$, where $L = F_n$ (a Fibonacci number), and where $N = 0,...,L$ (all possible densities in the fermionic case),   We would like to prove that the closure of $A$ is the set $[0,1]$.  This can easily be shown by demonstrating that for every $x \in [0,1]$ and every $\epsilon > 0 $ there exists an $a \in A$ which is in the open ball around $x$ defined by $(x-\epsilon,x+\epsilon)$.   It is easy to see that this is always possible, since the set of all densities $N/L$ is evenly spaced in the interval $[0,1]$, so one should choose $n$ large enough so that $1/F_n$ is smaller than $2 \epsilon$.   As $n$ is increased further one can form a sequence which converges to $x$.


\begin{thebibliography}{9}


\bibitem{Abrahams79} E. Abrahams, P. W. Anderson, D. C. Licciardello, and T. V. Ramakrishnan, "Scaling Theory of Localization: Absence of Quantum Diffusion in Two Dimensions" {\it Phys. Rev. Lett.} {\bf 42} 673 (1979).

\bibitem{Langedijk09} A. Langedijk, B. van Tiggelen, and D. S. Wiersma, "Fifty years of Anderson localization" {\it Physics Today} {\bf 62} 24 (2009).

\bibitem{Evers08} F. Evers and A. D. Mirlin, "Anderson transitions" {\it Rev. Mod. Phys.} {\bf 80} 1355 (2008).
 
\bibitem{Aubry80} S. Aubry and G. Andr\'{e}, "Analyticity breaking and Anderson localization in incommensurate lattices." {\it Ann. Isr. Phys. Soc.} {\bf 3} 133 (1980).

\bibitem{Martinez18} A. J. Martinez, M. A. Porter, and P. T. Kevrekidis, "Quasiperiodic granular chains and Hofstadter butterflies" {\it Philos. Trans. A} {\bf 376} 20170139 (2018). 

\bibitem{Dominguez-Castro19} G. A. Dominguez-Castro, R. Paredes,  "The Aubry–André model as a hobbyhorse for understanding the localization phenomenon", {\it Eur. J. Phys.} {\bf 40} 045403 (2019).

 \bibitem{Harper55} P. G. Harper, "Single Band Motion of Conduction Electrons in a Uniform Magnetic Field"{\it Proc. Phys. Soc. A} {\bf 68} 874 (1955).
 
 \bibitem{vonKlitzing80} K. von Klitzing, G. Dorda, and M. Pepper, "New Method for High-Accuracy Determination of the Fine-Structure Constant Based on Quantized Hall Resistance" {\it Phys. Rev. Lett.} {\bf 45} 494 (1980).
 
 \bibitem{Tsui82} D. C. Tsui, H. L. Stormer, and A. C. Gossard, "Two-Dimensional Magnetotransport in the Extreme Quantum Limit"{\it Phys. Rev. Lett.} {\bf 48} 1559 (1982).
 
  \bibitem{Thouless82} D. J. Thouless, M. Kohmoto, M. P. Nightingale, M. den Nijs, "Quantized Hall Conductance in a Two-Dimensional Periodic Potential" {\it Phys. Rev. Lett.} {\bf 49 } 405 (1982).

\bibitem{Jitomirskaya99} S. Ya. Jitomirskaya,  "Metal-Insulator Transition for the Almost Mathieu Operator" {\it Ann. Math.} {\bf 150}  1159 (1999).

\bibitem{Avila06} A. Avila, S. Jitomirskaya, "Solving the Ten Martini Problem"  In: J. Asch, A. Joye,  (eds) {\it Mathematical Physics of Quantum Mechanics}, Lecture Notes in Physics, vol 690. Springer, Berlin, Heidelberg . 

\bibitem{Avila09} A. Avila and S. Jitomirskaya "The Ten Martini Problem" {\it Ann. Math.} {\bf 170} 303 (2009).

\bibitem{Avila23} A. Avila, J. You, and Q. Zhou, "Dry Ten Martini Problem in the non-critical case" arxiv:2306.16254. 

\bibitem{Billy08} J. Billy, V. Josse, Z. Zuo, A. Bernard, B. Hambrecht, P. Logan, D. Cl\'{e}ment, L. Sanchez-Palencia, P. Bouyer, and A. Aspect, "Direct observation of Anderson localization of matter-waves in a controlled disorder" {\it Nature (London)} {\bf 453} 891 (2008).

\bibitem{Roati08} G. Roati, C. D'Errico, L. Fallani, M. Fattori, C. Fort, M. Zaccanti, G. Modugno, M. Modugno, and M. Inguscio, . "Anderson localization of a non-interacting Bose–Einstein condensate" {\it Nature (London)} {\bf 453} 895 (2008).

\bibitem{Modugno09} M. Modugno, "Exponential localization in one-dimensional quasi-periodic optical lattices" {\it New. J. Phys.} {\bf 11} 033023 (2009).

\bibitem{Kohlert19} T. Kohlert, S. Scherg, X. Li, H. P. L\"{u}schen, S. Das Sarma, I. Bloch, M. Aidelsburger, "Observation of Many-Body Localization in a One-Dimensional System with a Single-Particle Mobility Edge" {\it Phys. Rev. Lett. } {\bf 122} 170403 (2019).

\bibitem{Johansson91} M. Johansson and R. Riklund, "Self-dual model for one-dimensional incommensurate crystals including next-nearest-neighbor hopping, and its relation to the Hofstadter model" {\it Phys. Rev. B} {\bf 43} 13468 (1991).
  
\bibitem{Biddle10} J. Biddle and S. Das Sarma, "Predicted Mobility Edges in One-Dimensional Incommensurate Optical Lattices: An Exactly Solvable Model of Anderson Localization"  {\it Phys. Rev. Lett.} {\bf 104} 070601 (2010).

 \bibitem{Biddle11}  J. Biddle, D. J. Priour Jr., B. Wang, and S. Das Sarma, "Localization in one-dimensional lattices with non-nearest-neighbor hopping: Generalized Anderson and Aubry-André models" {\it Phys. Rev. B} {\bf 83} 075105 (2011).
  
 \bibitem{Ganeshan13} S. Ganeshan, K. Sun, and S. Das Sarma, "Topological Zero-Energy Modes in Gapless Commensurate Aubry-André-Harper Models"  {\it Phys. Rev. Lett.} {\bf 110} 180403 (2013).
 
   \bibitem{Ganeshan15} S. Ganeshan, J. H. Pixley, and S. Das Sarma, "Nearest Neighbor Tight Binding Models with an Exact Mobility Edge in One Dimension" {\it Phys. Rev. Lett.} {\bf 114} 146601 (2015).
 
 \bibitem{Bistritzer11} R. Bistritzer and A. H. Macdonald, "Moiré bands in twisted double-layer graphene", {\it Proc. Nat. Acad. Sci. USA} {\bf 108} 12233 (2011).

\bibitem{Monthus17} C. Monthus, "Multifractality in the generalized Aubry-Andre quasiperiodic localization
model with power-law hoppings or power-law Fourier coefficients" {\it Fractals} {\bf 27} 1950007 (2017).

 \bibitem{Li20} X. Li and S. Das Sarma, "Mobility edge and intermediate phase in one-dimensional incommensurate lattice potentials", {\it Phys. Rev. B} {\bf 101} 064203 (2020).

\bibitem{Padhan22} A. Padhan, M. K. Giri, S. Mondal, and T. Mishra, "Emergence of multiple localization transitions in a one-dimensional quasiperiodic lattice" {\it Phys. Rev. B} {\bf 105} L220201 (2022).
 
\bibitem{Goncalves23a} M. Gon\c{c}alves, B. Amorim, E. V. Castro, and P. Ribeiro, "Critical Phase Dualities in 1D Exactly Solvable Quasiperiodic Models" {\it Phys. Rev. Lett.} {\bf 131} 186303 (2023).

\bibitem{Goncalves23b} M. Gon\c{c}alves, B. Amorim, E. V. Castro, and P. Ribeiro, "Renormalization group theory of one-dimensional quasiperiodic lattice models with commensurate approximants" {\it Phys. Rev. B} {\bf 108} L100201 (2023).

\bibitem{Dziurawiec24} M. Dziurawiec, J. O. de Almeida, M. L. Bera, M. P\/{l}odzie\'{n}, M. M. Maśka, M. Lewenstein, T. Grass, and U. Bhattacharya, "Unraveling multifractality and mobility edges in quasiperiodic Aubry-André-Harper chains through high-harmonic generation" {\it Phys. Rev. B} {\bf 110} 014209 (2024).

\bibitem{Chi24} R. Chi, J. J. Yu, C. Murthy, and T. Xiang, "Luttinger liquid phase in the Aubry-André Hubbard chain", {\it Phys. Rev. B} {\bf 110} 165117 (2024).
 
 \bibitem{Zhang25} Z. H. Zhang, H. C. Kou, P. Li, "Critical dynamics and its interferometry in the one-dimensional $p$-wave-paired Aubry-André-Harper model", {\it Phys. Rev. B} {\bf 112} 014310 (2025).
 
 \bibitem{Goswami25} A. Goswami, P. Chatterjee, R. Modak, and S. Sahoo, "Subsystem localization in a two-leg ladder system", {\it Phys. Rev. B} {\bf 112} 144205 (2025).
 
 \bibitem{Lu25} F Lu, A. Zhou, S. Cheng, and G. Xianlong, "Wigner distribution, Wigner entropy, and anomalous transport of a generalized Aubry-André model", {\it Phys. Rev. B} {\bf 112} 174206 (2025).
 
 \bibitem{Gandhi25} S. Gandhi, J. N. Bandyobadhyay, "Superconducting $p$-wave pairing effects on one-dimensional non-Hermitian quasicrystals with power law hopping", {\it Phys. Rev. B} {\bf 111} 174210 (2025).
 
 \bibitem{Sahoo26} A. Sahoo and D. Rakshit, "Enhanced sensing of a weak Stark field under the influence of Aubry-André-Harper criticality" {\it Phys. Rev. B} {\bf 113} 022601 (2026).
 
 \bibitem{Zhang26} C. Zhang, "Interaction-induced quasicrystalline order: Emergence of quasisolid and quasisupersolid phases", {\it Phys. Rev. B} {\bf 113} L220502 (2026).
 
 \bibitem{Liu26} T. Liu, "Dual-spaces invariance as a concise criterion for multifractal critical states", {\it Int. J. Mod. Phys. B} {\bf 40} 2650093 (2026).
 
\bibitem{Bonsel26} F. Bönsel, F. K. Kunst, and F. Roccati,  "Fibonacci Waveguide Quantum Electrodynamics", Quantum {\bf 10} 2081 (2026).

\bibitem{Jeon26} J. Jeon and S. Sakai, "Quantum hyperuniformity and quantum weight", {\it Phys. Rev. B} {\bf 113} L241113 (2026).

\bibitem{Cookmeyer20} T. Cookmeyer, J. Motruk, J. E. Moore, "Critical properties of the ground-state localization-delocalization transition in the many-particle Aubry-André model", {\it Phys. Rev. B} {\bf 101}174203 (2020).

 \bibitem{Varma15} V. Kerala Varma and S. Pilati, "Kohn's localization in disordered fermionic systems with and without interactions" {\it Phys. Rev. B} {\bf 92} 134207 (2015).
 
 \bibitem{Hetenyi24} B. Het\'{e}nyi, "Scaling of the bulk polarization in extended and localized phases of a quasiperiodic model" {\it Phys. Rev. B} {\bf 110} 124125 (2024).

 \bibitem{Hetenyi25} B. Het\'{e}nyi and I. Balogh, "Numerical study of the localization transition of Aubry-André type models" {\it Phys. Rev. B} {\bf 112} 144203 (2025).
 
 \bibitem{Mastropietro15} V. Mastropietro, "Localization of Interacting Fermions in the Aubry-André Model" {\it Phys. Rev. Lett.} {\bf 115} 180401 (2015).

 \bibitem{Xu19} S. Xu, X. Li, Y.-T. Hsu, B. Swingle, S. Das Sarma, "Butterfly effect in interacting Aubry-Andre model: Thermalization, slow scrambling, and many-body localization", {\it Phys. Rev. Research } {\bf 1}  032039(R) (2019).
 
 \bibitem{Huang24} K. Huang, D. Vu, S. Das Sarma, X. Li, "Interaction-enhanced many body localization in a generalized Aubry-Andre model" {\it Phys. Rev. Res.} {\bf 6}, L022054 (2024).

 \bibitem{Bergman57} G. Bergman, "A Number System with an Irrational Base" {\it Mathematics Magazine} {\bf 31} 98 (1957).
 
   \bibitem{Resta98} R. Resta, "Quantum-Mechanical Position Operator in Extended Systems" {\it Phys. Rev. Lett.} {\bf 80} 1800 (1998).
  
  \bibitem{Resta99} R. Resta and S. Sorella, "Electron Localization in the Insulating State", {\it Phys. Rev. Lett.} {\bf 82} 370 (1999). 

 \bibitem{Vajda89} S. Vajda, "Fibonacci and Lucas Numbers, and the Golden Section: Theory and Applications", Ellis Horwood Limited, Chichester, U.K., 1989.

 \bibitem{Zeckendorf72} E. Zeckendorf, "Représentation des nombres naturels par une somme de nombres de Fibonacci ou de nombres de Lucas", {\it Bull. Soc. R. Sci. Liège} {\bf 41} 179 (1972).
 
 \bibitem{Samons94}  J. D. Samons, "A Relationship Between the Fibonacci Sequence and Cantor's Ternary Set", UNF Graduate Theses and Dissertations. 285, (1994).  (https://digitalcommons.unf.edu/etd/285).
 
 
 
 







\end{thebibliography}
\end{document}